\documentclass[twocolumn]{revtex4-1}

\usepackage{graphicx}
\usepackage{relsize}

\begin{document}

\title{An orientation-field model for polycristalline solidification with a singular coupling between order and orientation}

\author{Herv\'e Henry}
\affiliation{Physique de la Mati\`ere Condens\'ee, \'Ecole Polytechnique, CNRS,
91128 Palaiseau, France}
\author{Jesper Mellenthin}
\affiliation{Physique de la Mati\`ere Condens\'ee, \'Ecole Polytechnique, CNRS,
91128 Palaiseau, France}
\author{Mathis Plapp}
\affiliation{Physique de la Mati\`ere Condens\'ee, \'Ecole Polytechnique, CNRS,
91128 Palaiseau, France}

\date{\today}

\begin{abstract}
The solidification of polycrystalline materials can be modelled by
orientation-field models, which are formulated in terms of two
continuous fields: a phase field that describes the thermodynamic
state and an orientation field that indicates the local direction of the
crystallographic axes. The free-energy functionals of existing models generally 
contain a term proportional to the modulus of the orientation gradient,
which complicates their mathematical analysis and induces artificial 
long-range interactions between grain boundaries. We present an
alternative model, in which only the square of the orientation gradient
appears, but in which the phase and orientation fields are coupled by 
a singular function that diverges in the solid phase. We show that 
this model exhibits stable grain boundaries whose interactions 
decay exponentially with their distance. Furthermore, we demonstrate 
that the anisotropy of the surface energy can be included
while preserving the variational structure of the model. Illustrative
numerical simulations of two-dimensional examples are also presented.
\end{abstract}

\maketitle

\section{Introduction}

In polycristalline materials that consist of multiple crystalline 
grains of the same thermodynamic phase, the grain size and texture of 
the material largely determine mechanical properties such as toughness 
and yield stress. Therefore, a large effort has been devoted to
understanding the relation between the processing conditions
during solidification and subsequent ageing of a material and
the resulting grain structure. From a fundamental viewpoint,
this is an example of self-organization \cite{Cross1993} : a complex 
structure emerges from a structureless initial state (the liquid), 
and the resulting network of grain boundaries evolves with time
and coarsens \cite{Atkinson88,Thompson00}.

Despite a large body of work, many questions regarding this
self-organization process remain open. As often in such complex
problems, a comparison between experiments and numerical modeling
could be highly useful. Therefore, it is desirable to develop 
quantitative modeling tools for the solidification and evolution 
of polycristalline materials. Among the many modelling approaches, 
the phase-field method
has rapidly gained in popularity in recent years due to its versatility 
and robustness. For recent reviews on its applications in materials
science, see Refs. \cite{Boettinger02,Chen2002,Steinbach09}.

Numerous phase-field models for the description of polycristalline
solidification and grain growth have already been developed. They 
can be grouped in two categories. In the so-called multi-phase-field 
approach \cite{multiphase1,Fan97,nestler2005,Kim06,Moelans08}, each grain 
is described by an individual phase field. The grain boundary energies 
are determined by the coupling between the phase fields. Despite the necessity
to handle  a large number of phase fields,
this approach can be made quite efficient from a numerical point of 
view \cite{Kim06,Vanherpe07}. Nevertheless, it suffers from the fact that 
for each simulation the coupling coefficients between the phase fields 
need to be computed (and stored) in order to reproduce the proper 
grain boundary energies \cite{Moelans08}. Moreover, since the
crystalline orientation in each grain is fixed, phenomena such as
grain rotation and plastic deformation cannot easily be described.
Finally, from a fundamental point of view, a large number of fields
certainly does not constitute the most ``economic'' description of
a polycristalline structure, since the state of matter in any
space point can be specified by a set of a few variables, such as 
the degree of order and the local lattice orientation.

As a consequence of the latter consideration, alternative models
have been developed that use a single phase field coupled to an orientation 
field \cite{warren98,Kobayashi00,Lobkovsky02,Warren03,Granasy04,Tang06}. 
The phase field describes the local thermodynamic state -- either 
solid ($\phi=1$) or liquid ($\phi=0$) -- and the orientation field 
is used to describe the local orientation of the crystallographic 
axes in the solid with respect to some fixed reference system.
In two dimensions, the orientation field is simply 
an angle $\theta({\mathbf x},t)$, while in three dimension a 
more complex formalism is needed \cite{kobayashi2005,Pusztai2005}. 
In this framework, a grain boundary is a diffuse but localized region 
in space between two regions of the same thermodynamic state (both are 
solid) but different crystalline orientations. The change of the
crystalline orientation is concentrated in the grain boundary where the 
phase field is no longer equal to $1$. 

While this kind of description is appealing, it turns out that the
construction of a viable model is far from simple because of the
need to obtain localized grain boundaries. Let us consider the 
two-dimensional (2D)
case with a scalar angle field and try to construct a free energy
functional that leads to localized solid-liquid interfaces and
grain boundaries. For the solid-liquid interfaces, this is
straightforward: since there are two privileged and distinct 
thermodynamic states (solid and liquid), diffuse interfaces
arise from the free energy balance between a double-well potential 
and a square-gradient term for the phase field. In the case of 
grain boundaries, matters are not so simple. Indeed, there is no
privileged angle and the free energy must be rotationally invariant. 
Therefore, a potential that depends on the angle is forbidden, and
only terms that contain spatial derivatives of $\theta$ can be used  
in the free energy. Then, in order to model localized grain
boundaries, at a first glance it seems sufficient to introduce  
a term that couples the energy
cost of angular variations to the local value of the phase 
field and induces a sufficiently high energy penalty for angle
variations in the solid.

However, a simple scaling argument, reviewed in detail below, 
shows that localized grain boundaries cannot be obtained by
a straightforward coupling of a square-gradient term in $\theta$
to the phase field. In fact, such a model is equivalent to the
usual thermodynamic description of nematic liquid 
crystals \cite{deGennes}, in which indeed no ``grain boundaries''
(localized changes in the director field)  exist.
The solution proposed in the previous
orientation-field models is to include an additional term
proportional to $|\nabla \theta|$ in the free energy. Localized
grain boundaries then arise from the balance between this term
and the square gradient in $\theta$. While this approach
yields localized grain boundaries, it also has several problems.
First, from a fundamental viewpoint it seems difficult to
justify such a form in the spirit of a Landau expansion. Second,
the functional derivative of this non-analytic term leads to
a singular diffusion equation for $\theta$ and to non-local 
interactions between grain boundaries \cite{Kobayashi99}. The latter 
effect can be removed by a proper regularization \cite{Warren03}, but 
this introduces additional parameters into the model.

Here, we propose an alternative approach. Our model contains only 
a square gradient term in $\theta$, but relies on the use of a 
singular coupling function between orientation and phase field to 
make the energy cost of angular variations in the solid effectively 
infinite. Similar ideas were proposed some years ago by Lusk \cite{Lusk99}, 
but were to our best knowledge not pursued any further until now.
The interesting point about this approach is that the
equations for the equilibrium grain boundaries and solid-liquid
interfaces are ordinary (non-singular) differential equations
that can be analyzed using standard mathematical tools and concepts 
from dynamical systems theory. Furthermore, it turns out that the
dynamical equations for the angle field show the proper behavior,
that is, no long-range interactions between grain boundaries 
arise. Therefore, we believe that this approach can be a useful
alternative to existing orientation-field models.

The remainder of the paper is organized as follows. In 
Sec.~\ref{sec_model}, we first outline the model (for isotropic
interfaces) and then show how to determine a suitable coupling 
function between phase and orientation fields that yields 
localized grain boundaries. Furthermore, we discuss in detail 
the properties of these grain boundary solutions. In Sec.~\ref{sec_anis},
we then extend the model to include interfacial anisotropy, and
present numerical tests both in one and two dimensions to show 
that the model has the desired properties. Perspectives for future
work are discussed in Sec.~\ref{sec_end}.

\section{Model}
\label{sec_model}
\subsection{General remarks and outline}

We consider a two-dimensional system in which the crystal orientation
can be described by a single angle field $\theta({\mathbf x},t)$; the
local state of the system is described by a phase field $\phi({\mathbf x},t)$.
We start from a free energy of the form
\begin{equation}
F=\int d{\mathbf x} \left[ \frac{D}{2}\nabla \phi)^2+H V(\phi,T)+ K
g(\phi)(\nabla \theta)^2\right],
\end{equation}
where $D$, $H$ et $K$ are constants. Furthermore, the dimensionless
function $V(\phi,T)$ is a temperature-dependent double-well 
potential with minima at $\phi=1$ (solid) and $\phi=0$ (liquid), 
and $g$ is a function that tends to zero for $\phi \to 0$. For 
simplicity, we have neglected crystalline anisotropy for the moment; 
its inclusion will be discussed below.

Let us first consider a coupling function $g(\phi)$ that takes
a finite value $g(1)$ in the solid. The scaling argument which 
proves that there are no localized grain boundary solutions
in this case is as follows. In the absence of anisotropy, the 
grain boundary solution must depend only on the total angle 
change (misorientation) $\Delta\theta$ between the two grains. For a 
solid of size $L$ with a homogeneous continuous orientation variation 
between $\Delta \theta/2$ and $-\Delta \theta/2$, the energy 
cost (with respect to a state of constant $\theta$) is 
$g(1) KL (\Delta \theta/L)^2$ which decreases monotonously with
growing $L$ and tends to zero when $L\to \infty$. 
In contrast, any localized grain boundary
solution has an energy that is independent of the system size
(and includes the energy cost for the variation of the phase
field). For sufficiently large $L$, the uniform angle variation
has therefore always lower energy than a localized grain boundary.
More precisely, whereas grain boundaries can be stable in 
finite systems depending on the values of $L$ and $g(1)$, 
an isolated grain boundary in an infinite system is always 
unstable. This scaling argument also clarifies 
why a term of the form $|\nabla\theta|$ cures this problem:
with such a term, the energy of a homogeneous variation of the
angle is independent of the system size and can be made larger
than the energy of a localized solution by a proper choice of
coupling functions and parameters.

Here, we propose an alternative approach to achieve the same goal. 
Since the above argument holds only for a finite value of $g(1)$, 
a possible  way out is to use a {\em singular} coupling function 
$g(\phi)$ that tends to infinity when $\phi\to 1$. The rationale 
for this idea is that in a crystalline material any continuous 
variation of the local orientation implies an elastic or plastic 
strain, which has a high energy cost. We will show below that we
can indeed choose a singular coupling function that leads to grain
boundaries with the desired properties. This function cannot be 
explicitly linked to microscopic physics or an elastic model; 
our approach is therefore essentially phenomenological. 

For the sake of definiteness, we use a coupling function of the form
\begin{equation}
g(\phi)=\frac{f(\phi)}{(1-\phi)^\alpha},
\end{equation}
where $\alpha$ is some positive real exponent and $f$ is a 
polynomial in $\phi$ which satisfies $f(0)=0$ and $f(1)=1$.

It is useful for the following developments to non-dimensionalize
the free energy functional and the involved length scales. The
functional has three parameters: $D$ and $K$ have dimensions of
energy per unit length, and $H$ sets a free energy (density) scale 
for the (dimensionless) double-well potential $V(\phi,T)$. 
Dividing the whole functional by $H$, we obtain
\begin{equation}
F=\int d{\mathbf x} \left[\frac{W_\phi^2}{2}(\nabla \phi)^2+ V(\phi,T)+ 
W_\theta^2 g(\phi)(\nabla \theta)^2\right],
\end{equation}
where $W_\phi=\sqrt{D/H}$ and $W_\theta=\sqrt{K/H}$ are the
characteristic length scales associated with solid-liquid
interfaces and grain boundaries, respectively. Scaling all
lengths with $W_\phi$ and defining $\mu=W_\theta/W_\phi$,
we finally obtain
\begin{equation}
F=\int d{\mathbf x} \left[\frac{1}{2}(\nabla \phi)^2+ V(\phi,T)+ 
\mu^2 g(\phi)(\nabla \theta)^2\right].
\label{general_form_F}
\end{equation}
In the following, we consider the case $\mu=1$ that can be obtained 
by a proper rescaling of $\theta$ by $\mu$. It should be pointed
out that this is also the best case for a useful model since then
the characteristic thickness of solid-liquid interfaces and grain
boundaries are comparable.

The equations of motion for the two fields are obtained by the
standard variational procedure
 \cite{warren98,Kobayashi00,Lobkovsky02,Warren03,Granasy04,Tang06},
\begin{eqnarray}
\label{phi_evolution}
P(\phi,\nabla\theta)\tau_\phi \partial_t \phi & = & - \frac{\delta F}{\delta \phi}, \\
\label{theta_evolution}
Q(\phi,\nabla\theta)\tau_\theta \partial_t \theta & = & - \frac{\delta F}{\delta \theta},
\end{eqnarray}
where $\tau_\phi$ and $\tau_\theta$ are relaxation times, and
$P(\phi,\nabla\theta)$ and $Q(\phi,\nabla\theta)$ are 
(dimensionless) mobility functions.

For the solidification of a pure substance, the motion of solid-liquid
interfaces is linked to heat transport by the Stefan condition (the
latent heat released or consumed at moving interfaces must be transported
by the local diffusion currents). The model therefore has to be completed
by an equation for the temperature field. We define a dimensionless
temperature $u$ by
\begin{equation}
u(\mathbf{x},t) = \frac{T(\mathbf{x},t)-T_m}{L/c},
\end{equation}
where $T_m$, $L$, and $c$ are the melting temperature, the latent heat
of fusion, and the specific heat, respectively. The field $u$ obeys
the same equation as in the standard phase-field model,
\begin{equation}
\partial_t u = D_{\rm th} \nabla^2 u + \partial_t \phi,
\label{uevolution}
\end{equation}
with $D_{\rm th}$ the thermal diffusion coefficient.

It turns out that in order to obtain a model with the desired
properties, a certain number of conditions on the exponent $\alpha$,
the polynomial $f$ and the double-well potential $V$ have to be
satisfied that are linked in a sometimes subtle way to the
internal structure of the grain boundary. In order to facilitate
the understanding of the following developments, it seems useful
at this point to state a summary of these results, while the 
details of the arguments will be given below. The main points 
are the following.
\begin{enumerate}
\item
Localized grain boundary solutions which smoothly connect to 
an infinite bulk solid on both sides can exist only for 
$\alpha\ge 2$.
\item
For a fixed total misorientation $\Delta\theta$ and increasing
system size $L$, the energy cost of a uniformly strained solid 
tends to zero for $\alpha<2$ and to infinity for $\alpha >2$. 
Together with condition 1, 
this yields that the only reasonable value for $\alpha$ is $2$.
\item
The requirement that the angle variation is strongly concentrated
in the center of the grain boundary imposes that the first term
appearing in the polynomial $f$ is of order $\phi^3$.
\item
A uniformly strained solid must be stable against the spontaneous
formation of grain boundaries. This imposes the relation $ab-d>0$,
where $a=V''(1)/2$, $d=-V'''(1)/6$, and $b=-f'(1)$, and primes
denote a derivative with respect to $\phi$.
\item
For $\alpha=2$, the competition between the gradient terms of
the phase field and the orientation field makes it impossible
to obtain stable grain boundaries with misorientations smaller
than a critical misorientation $\Delta\theta_m$. The magnitude
of $\Delta\theta_m$ is controlled by the same combination of
parameters that appears in condition 4, $ab-d$.
\end{enumerate}

In particular, the last point has implications for the choice
of the double-well potential $V$. In the standard phase-field
model for solidification, this potential is
$V(\phi)=\phi^2(1-\phi)^2+u\lambda(10\phi^3-15\phi^4+6\phi^5)$, 
where $\lambda$ is a dimensionless coupling 
parameter. For this potential, the second derivatives with
respect to the phase field, $V''(\phi,T)$, are independent 
of temperature in both wells ($\phi=0$ and $\phi=1$); however, 
the third derivatives depend on the temperature. For our model,
this would yield a temperature-dependent minimal grain boundary
misorientation. We prefer to use a potential where, in addition
to the above properties, the third derivative of $V$ in the
solid well is kept constant. Furthermore, since for the calculations 
of the grain boundary energy the solid state is the reference state,
we wish to keep the energy of the solid potential well independent
of temperature and equal to zero, whereas the height of the
liquid potential minimum varies with temperature.

A set of model functions which satisfies all the above
requirements and conditions is
\begin{equation}
g=\frac{7\phi^3-6\phi^4}{(1-\phi)^2}
\label{gfunc}
\end{equation}
\begin{equation}
V=\phi^2(1-\phi)^2-u\lambda(1-20\phi^3+45\phi^4-36\phi^5+10\phi^6)
\label{vfunc}
\end{equation}
These functions will be used in the numerical examples given below.
In the following, we will first give the details of the arguments
that lead to the restrictions on these functions, and then discuss 
their consequences on the structure of the grain boundary solutions,
in particular concerning the dependence of these solutions on 
temperature.

\subsection{Construction of the coupling function: the details}

\begin{figure}
\centerline{
\includegraphics[width=0.5\textwidth]{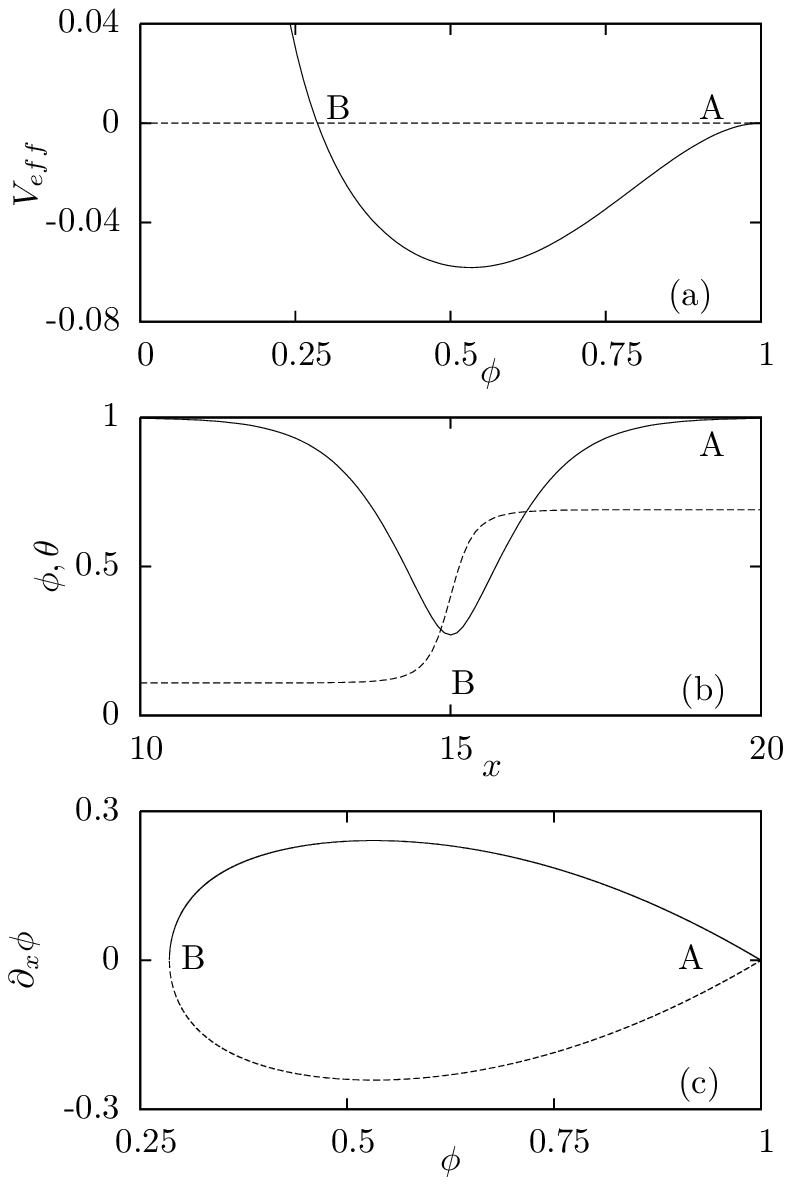}}
\caption{\label{fig_veff} (a) Schematic drawing of a typical 
effective potential $V_{\rm eff}$. The starting point of the 
trajectory corresponding to $\phi=1$ is A, the turning point is B.
In both A and B, $\partial_x \phi =0$. (b) Typical profiles of
$\phi$ (solid line) and $\theta$ (dashed line) through a grain
boundary. A corresponds to the limit $\phi \to 1$ and $x\to \pm\infty$. 
(c) Grain boundary solution, plotted in the phase space spanned 
by $\phi$ and $\partial_x\phi$.} 
\end{figure}

Our model should have one-dimensional grain boundaries as 
equilibrium solutions. In the framework used here, a grain boundary 
is a localized interface (region where $\phi$ is smaller than 1)
between two solids (semi-infinite regions where $\phi=1$) with 
different crystal orientations. Therefore, we are looking for 
appropriate stationary solution of Eqs. (\ref{phi_evolution})
and (\ref{theta_evolution}). The calculation of the functional
derivative yields
\begin{eqnarray} 
0&=&\partial_x(g(\phi)\partial_x \theta),\label{eq_angle}\\
0&=&-V'(\phi,T)+\partial_{xx}\phi-g'(\phi)(\partial_x \theta)^2.\label{eq_phi}
\end{eqnarray} 
From Eq.~(\ref{eq_angle}), we obtain that at equilibrium
\begin{equation}
g(\phi)\partial_x\theta=C,
\label{eq_thetaprime}
\end{equation} 
where $C$ is a constant to be determined later. Inserting this 
expression into Eq.~(\ref{eq_phi}) yields 
\begin{eqnarray}
\partial_{xx}\phi&=&V'(\phi,T)+g'(\phi)\frac{C^2}{g(\phi)^2},\\
&=&-\frac{d}{d\phi}\left[-V(\phi,T)+\frac{C^2}{g(\phi)}\right]\label{eq_phi_eff}.
\end{eqnarray}
The second equality puts this equation in a form suitable for the
use of the well-known analogy with a mechanical system: replacing 
$x$ by $t$ and $\phi$ by $y$, one obtains the equation of a particle
with coordinate $y$ moving in a potential 
\begin{equation}
V_{\rm eff}(y)=-V(y,T)+\frac{C^2}{g(y)}
\label{veffdef}
\end{equation} 
(see Fig.~\ref{fig_veff} (a)). The grain boundary solution 
corresponds then to the trajectory of a particle that starts
with zero velocity at the unstable equilibrium point $y=1$, 
(the point A in Fig.~\ref{fig_veff}(a), which corresponds 
to the solid state) and reaches the turning point $y=y_0$ 
(the point B in Fig.~\ref{fig_veff} (a)), in an infinite 
time and comes back. The infinite 
duration of the trajectory corresponds to the condition that the
limit of $\phi$ in the bulk is $\phi=1$ (solid) on both sides of the
grain boundary. This leads to the requirement that $\phi=1$
must be an equilibrium point of $V_{\rm eff}$, i.e.
\begin{eqnarray}
\left. \frac{d V_{\rm eff}}{d\phi}\right|_{\phi=1}
&=& -V'(1)-C^2(1-\phi)^{\alpha-1}\frac{\alpha f + (1-\phi)f'}{f^2}=0,
\end{eqnarray}
where we have supposed $\alpha>1$ for $V_{\rm eff}$ to be differentiable
in $\phi=1$.
In addition, $\phi=1$ must correspond to an unstable equilibrium 
point of $V_{\rm eff}$, that is, the second derivative of the effective
potential must be negative for $\phi\to 1$. A simple argument
helps to determine the sign of the second derivative. Since the
double-well potential $V(\phi,T)$ has a minimum at $\phi=1$, we
can expand it around this value as 
$V(\phi,T)\approx V(1,T)+V''(1,T)(1-\phi)^2/2$.
Then the effective potential becomes 
\begin{equation}
V_{\rm eff}=-V(1,T)+(1-\phi)^2\left(-V''(1,T)/2+C^2\frac{(1-\phi)^{\alpha-2}}{f(\phi)}\right).
\end{equation}
From this expression, it is clear that the sign of $V''_{\rm eff}$ 
at $\phi=1$ is determined by the sign of the function
$s(\phi)=(-V''(1,T)/2+C^2(1-\phi)^{\alpha-2}/f(\phi))$ at $\phi=1$. 
If $\alpha-2>0$, the second term in this function tends to zero
for $\phi\to 1$, whereas $-V''(1,T)$ is finite and negative. 
Therefore, $s(1)$ is negative. In contrast, if $\alpha-2<0$, the
second term becomes dominant for $\phi\to 1$, and $s$ is positive.
If $\alpha=2$, $s=(-V''(1,T)/2+C^2/f(\phi))$. In this case, since 
$f(1)=1$, $s$ changes sign when $C^2$ crosses $-V''(1,T)/2$. 

In summary, the fact that $1$ must be an unstable equilibrium point of
$V_{\rm eff}$ imposes  to choose $\alpha \geq 2$. We proceed by showing
that the study of the homogeneous solid solution imposes $\alpha=2$. 
To do this, we consider a long homogeneous solid of length $L$ where 
$\theta$ is varying from $\Delta\theta/2$ to  $-\Delta \theta/2$ with
constant gradient $\partial_x\theta=\Delta\theta/L$. The scaling argument
outlined above shows that a regular coupling function yields an 
energy cost of such a homogeneous solution that goes to zero 
when $L\to\infty$ at constant $\Delta\theta$; 
we would hence like to construct a model in which this is not the case.
Qualitatively, it is easy to see why this can be achieved with a
singular coupling function. As long as $\partial_x\theta$ is
different from zero, the singularity forbids the phase field to
take the value $\phi=1$ corresponding to the solid, since this
would imply an infinite energy cost. However, when $\partial_x\theta$
decreases, the phase field can get closer and closer to $\phi=1$.
Therefore, whereas $(\partial_x\theta)^2$ decreases, the value
of $g(\phi)$ increases, which can be used to tune the dependence
of the total energy on the system size. The requirement that the
energy must remain finite (it should not go to zero, but not tend
to infinity either when $L\to\infty$) then fixes the exponent 
$\alpha$.

Let us now make this argument quantitative. Since we consider a 
weakly distorted solid, we can assume that $\phi$ is close to 1. 
Let $\delta=1-\phi$ where $\phi$ is the (constant) value of the 
phase field. Since $\phi$ is constant, the gradient in the phase
field does not contribute to the energy of this state. Keeping
the dominant term in $\delta$ in the two remaining terms of the
free energy functional, we obtain for the energy $E$ of the
state with homogeneous angle variation,
\begin{equation}
E=L\left[\frac{\delta^2}{2}V''(1)+\frac{(\partial_x\theta)^2}{\delta^\alpha}\right].
\label{eq_E_simp}
\end{equation}
Minimization of this expression with respect to $\delta$ yields
\begin{equation}
\delta=\left(\frac{\alpha(\partial_x\theta)^2}{V''(1)}\right)^{1/(\alpha+2)}.
\label{delta_min}
\end{equation}
Using this expression in Eq.~(\ref{eq_E_simp}), we find
\begin{equation}
E=L\left[Q\left(\frac{\Delta \theta}{L}\right)^{4/(\alpha+2)}\right],
\end{equation}
where $Q=[V''(1)/2][\alpha/V''(1)]^{2/(\alpha+2)}+[V''(1)/\alpha]^{\alpha/(\alpha+2)}$ 
is positive. The only possible value for $\alpha$ that yields a finite
energy in the limit of large $L$ is $\alpha=2$. With this choice, the 
energy of a large system with constant angle variation is proportional 
to $\Delta \theta$. 

The requirement that this homogeneous solution must be stable yields
another condition. To understand its origin, consider the curve
$E(L)$. If this curve is concave, there is an instability that
is analogous to spinodal decomposition: at fixed total $\Delta\theta$,
it is more advantageous for the system to concentrate the distortion
in one or several regions because this lowers the total energy; in
other words, there would be spontaneous formation of grain boundaries.
To prevent this, the curve $E(L)$ should be convex at least for low
values of the distortion (weakly strained solids), which implies
that it has to tend to the limit value from above. To find the
corresponding mathematical condition, the expansion in $\delta$
has to be pushed to higher order, where the energy writes (for $\alpha=2$)
\begin{equation}
E=L\left[a\delta^2+d\delta^3+\left(\frac{\Delta\theta}{L}\right)^2
\frac{1+b\delta+c\delta^2}{\delta^\alpha}\right]\label{eq_E_hom}
\end{equation}
where $a=V''(1)/2$, $d=-V'''(1)/6$, $b=-f'(1)$, and $c=f''(1)/2$.
Using the fact that $(\Delta \theta/L)^2\sim \delta^{\alpha+2}$
according to Eq. (\ref{delta_min}), it is straightforward to show 
that the requirement of $E(L)$ to be a decreasing function of $L$ 
for $L\to\infty$ implies
\begin{equation} 
ab+d >0\label{eq_1dhom_ineq_dev}.
\end{equation}

\begin{figure}
\centerline{
\includegraphics[width=0.5\textwidth]{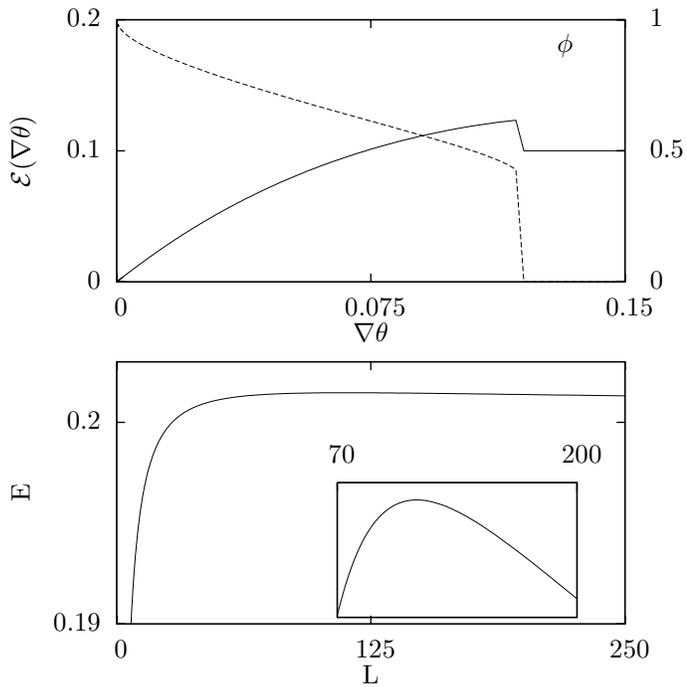}}
\caption{\label{fig_1dhom}(a) solid line: $E/L=\mathcal{E}(\Delta \theta)$ 
as a function of $\nabla\theta=\Delta \theta/L$, dashed line: $\phi$ as a 
function of $\Delta\theta/L$. For values of  $|\nabla \theta|>0.119$, the 
solid state with constant orientation gradient ceases to exist, 
and the only possible solution is the liquid state.
The undercooling is taken equal to 0.1 ($u=-0.1$). (b): curve $E=f(L)$ 
for $\Delta \theta=.1$ for the same parameters as in (a). Inset:
blowup of the vicinity of the maximum.}
\end{figure}

To finish with the study of the homogeneous solution, we find it useful to
present some numerical results using the complete model. The computation of
the energy as a function of $L$ for a fixed value of $\Delta \theta$ gives 
the expected result that the energy density depends only on the value of 
$\Delta\theta/L$~: $E/L=\mathcal{E}(\Delta\theta/L)$. The function
$\mathcal{E}(\Delta\theta/L)$ is plotted in figure \ref{fig_1dhom} (a).
It is interesting to note that in this case, the value of $dE/dL$ is a 
function of $\Delta\theta/L$ only. These results also show that, 
up to a rescaling of the energy $E$ and the length $L$, all the 
curves of the energy as a function of $L$ collapse onto a single master 
curve which is plotted in figure \ref{fig_1dhom} (b). This curve 
presents a maximum, the position of which is given by the solution
of the equation
\begin{equation}
\mathcal{E}(\Delta \theta/L)=\left(\frac{\Delta\theta}{L}\right)\mathcal{E}'(\Delta \theta/L).
\end{equation}

\subsection{Grain boundary solutions}

We now turn to a more complete description of the grain boundary equilibrium 
solutions for $\alpha\geq2$. They can be easily computed using the mechanical 
analog outlined above. For a given value of $C$, 
one first computes the value of $\phi_0$ corresponding 
to the turning point (B in fig. \ref{fig_veff}) by solving
$V_{\rm eff}(1)=V_{\rm eff}(\phi_0)$. Then, the equilibrium 
solution (going from the middle of the grain boundary to the solid 
state on one side of the grain boundary) corresponds to the 
trajectory of a moving particle in $V_{\rm eff}$ that starts at the
point B with zero velocity. Once $\phi(x)$ for the equilibrium solution has 
been computed, $\theta(x)$ is given by Eq.~(\ref{eq_angle}), and the total 
angle variation through the grain boundary can be obtained by simple 
integration. In order to compute the grain boundary solution for a 
given misorientation, the proper value of $C$ has to be found, for 
instance by using a Newton-Raphson method.

\begin{figure}
\centerline{
\includegraphics[width=0.5\textwidth]{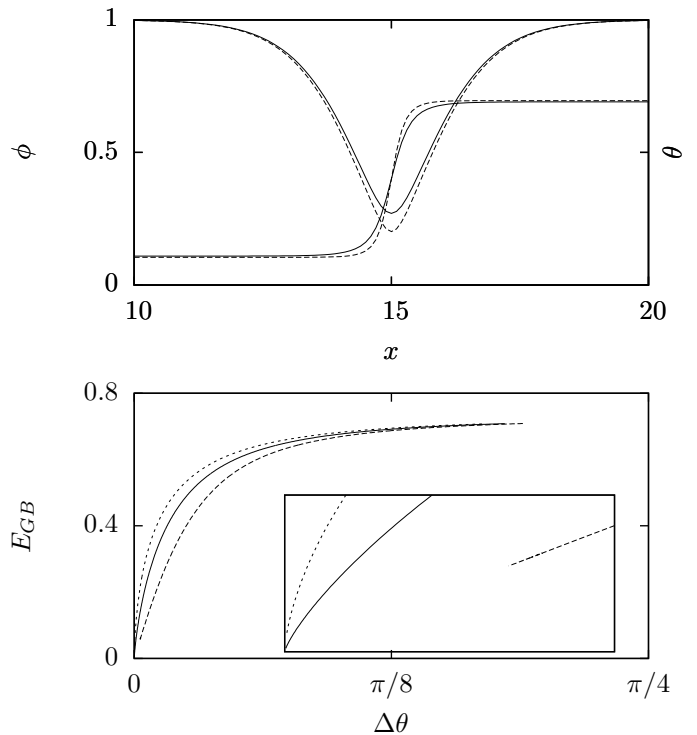}}
\caption{\label{fig_local1d}(a): typical grain boundary profiles for the same 
misorientation ($\Delta \theta=0.6$) and two values of $\alpha$: $\alpha=2$ (full lines)
and $\alpha=3$ (dashed lines). The dimensionless undercooling is 0.1.
(b): Grain boundary energy as a function of the misorientation $\Delta\theta$ 
for different values of $\alpha$ ($\alpha=2\mbox{, }\ 3\mbox{  and
} 4$). The inset displays a zoom around the origin and shows the singular 
behavior for $\alpha=3\mbox{ and }4 $ and
the fact that for $\alpha=2$ there is a small band of misorientations for which
no grain boundary solution exists. (Here $\mu^2=10$)}
\end{figure}

In figure \ref{fig_local1d}(a) we present a grain boundary solution for the 
same total angle variation $\Delta\theta=0.6$ for $\alpha=2$ and $\alpha=3$ 
(in the calculations for $\alpha=3$, we use the coupling function $g$ of
Eq.~(\ref{gfunc}), changing just the exponent). As predicted, the angle 
variation is concentrated in the vicinity of the minimum of $\phi$ (the turning 
point B), and is very small outside the region where $\phi$ differs from $1$.
This indicates that this is indeed a localized grain boundary solution.

Using the computed solutions, the grain boundary energy is given by
\begin{equation}
E_{GB}=\int dx  \left[\frac{1}{2}(\nabla \phi)^2+V(\phi,T)+ g(\phi)(\nabla
\theta)^2\right].
\end{equation}
(recall that we have chosen the homogeneous solid without any angle 
variation as the zero of the free energy density). We display this
energy as a function of the total angle variation $\Delta \theta$ 
in Fig.~\ref{fig_local1d} (b). 
In the cases  $\alpha=$ 3 and 4, for small values of $\Delta\theta$,
the energy has a power law behavior ($E_{GB}\sim (\Delta \theta)^\gamma$, with
$\gamma=$0.8 and 0.66, respectively). When $\Delta \theta$ increases
toward infinity (this is not absurd since we have previously rescaled
$\Delta \theta$ by $\mu$), the grain boundary energy converges 
toward a constant value.

In the case $\alpha=2$, the grain boundary energy also converges
toward a constant value when $\Delta\theta$ becomes large. However,
the behavior at small misorientations is very different: for 
small values of $\Delta \theta$, the grain boundary energy decreases 
linearly with the misorientation, but there exists a minimal value 
$\Delta\theta_m$ for which a grain boundary solution can exist (see inset). 
In fact, in the following we show that there is no grain boundary 
solution with angle variation smaller than this value in the case
$\alpha=2$.   	

\begin{figure}
\centerline{
\includegraphics[width=0.5\textwidth]{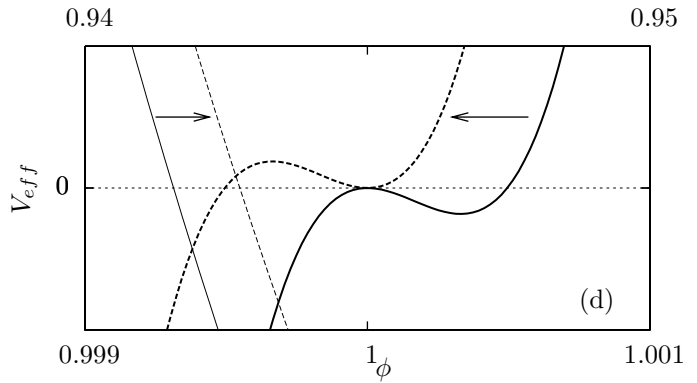}}
\caption{\label{fig_curvature}Effective potential $V_{\rm eff}$ for 
$\alpha=2$ and two values of $C$: $C=1\pm5\times 10^{-4}$ (dashed
and solid lines, respectively). The thin lines correspond to the 
interval $0.94<\phi<0.95$ (upper axis) and display the turning point
of the effective potential. The thick lines correspond to the 
interval $0.999<\phi< 1.001$ (lower axis) and display the vicinity
of the solid state. The arrows indicate the direction of change
when $C$ is lowered 
It can clearly be seen that the turning point
undergoes no qualitative change, whereas a new maximum appears close
to $\phi=1$ due to the fact that a zero of $V_{\rm eff}$ has crossed
the point $\phi=1$.
}
\end{figure}

As mentioned earlier, for $\alpha=2$, $\phi=1$ is not a maximum of 
the effective potential $V_{\rm eff}$ for all values of $C$. More 
precisely, $V_{\rm eff}$ can be formally written as
\begin{equation}
 V_{\rm eff}(\phi)=(1-\phi)^2[-p(\phi,T)+C^2/f(\phi)],
\end{equation}
where $p(T,\phi)$ is a function such that $p(\phi,T)(1-\phi)^2=V(\phi,T)$. 
Using the notations of Eq.~(\ref{eq_E_hom}), $V_{\rm eff}$ writes in the
vicinity of $\phi=1$:
\begin{equation}
V_{\rm eff}(\delta)=\delta^2
\left (
-a-d\delta+\frac{C^2}{1+b\delta+c\delta^2}
\right)
\end{equation}
with $\delta=1-\phi$. It can be seen that the curvature of $V_{\rm eff}$ 
changes sign in $\phi=1$ when $C^2$ becomes bigger than $a$. This means
that $\phi=1$ is now a local minimum of $V_{\rm eff}$. This change can
happen in two ways: either the effective potential becomes greater
that zero for all $\phi<1$ (the turning point B in Fig.~\ref{fig_veff}(a) 
moves to the right of the point A), or a new maximum develops between the
points A and B (this happens in our case, as illustrated in 
Fig.~\ref{fig_curvature}), which means that a zero of $V_{\rm eff}$ must 
cross the point A from above. It is straightforward to show that the latter 
always happens if $ab+d>0$. This explains the threshold value for 
$\Delta \theta$: grain boundary solutions with finite $\Delta\theta$ 
exist until the new maximum appears. Beyond this point, no grain boundary 
solution connecting to $\phi=1$ is possible any more.

We have also verified that all the grain boundary solutions found
up to now can be reached by simple integration of the equations of
motion, Eqs. (\ref{phi_evolution}) and (\ref{theta_evolution}),
at constant temperature (the field $u$ is held constant), when the
system is started from a homogeneous solid (with a value of the phase 
field slightly lower than unity) and a step function in the angle field.
The resulting stationary solutions as well as their energies are 
independent of the system size (for sufficiently large systems) and 
of the discretization used (for grid spacings $\Delta x$ smaller than 
about 0.5 $W$). This shows that the model formulation is robust,
and that the basin of attraction of the grain boundary solutions
is large enough to make the model suitable for simulations of
polycrystals.

\subsection{Temperature dependence of the grain boundary solutions}
 
In order to better characterize the model, we find it useful
to explore the temperature dependence of the grain boundary 
solutions. We start out with the regime that should be of most
practical interest, namely temperatures below the melting point
($T<T_m$, $u<0$).

\begin{figure}
\centerline{
\includegraphics[width=0.5\textwidth]{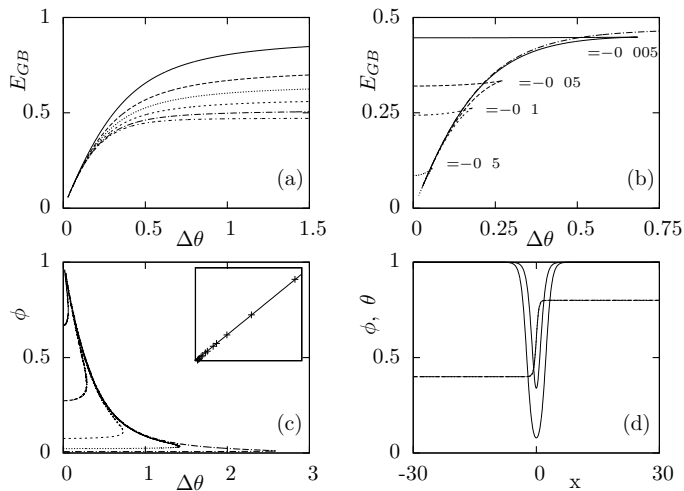}}
\caption{\label{Tdependance} \textbf{(a)~:} Grain boundary energy 
as a function of $\Delta \theta$ for different values of the undercooling.
The dash-dotted curve corresponds to $u=0$. 
\textbf{(b)~:}
Grain boundary energy as a function of misorientation for different
values of the undercooling for overheated grain boundaries ($u>0$, 
negative undercooling). Grain boundary solutions exist only for a
finite range of misorientations.
\textbf{(c)~:}  minimal value of $\phi$ in the grain boundary in 
the case of overheated grain boundaries, as a function of
$\Delta\theta$, inset: value of the maximal angle variation through a stable
grain boundary as a function of $u^{-1/4}$. 
\textbf{(d) :} $\phi$ (solid)   and $\theta$ (dashed) as a function of
$x$ along both a grain boundary and a one-dimensional liquid nucleus
for the same angle variation $\Delta\theta=0.4$ and $u=0.005$.}
\end{figure}

From the discussion of the preceding section, it is clear that the 
minimal grain boundary angle strongly depends on the behavior of 
$V(\phi)$ close to the solid state. Since we have chosen $V(\phi)$
such that the solid well is independent of temperature, the 
small-angle solutions should not be much affected by temperature 
changes. The same argument indicates that changes in undercooling 
(increasing undercooling) should only appreciably modify the 
high-angle grain boundaries and their energies. Simulation results 
show that this is indeed the case (see Fig.~\ref{Tdependance}). As 
expected, for the double-well function of Eq.~(\ref{vfunc}) that 
was designed to maintain the combination $ab-d$ constant
(see appendix A), the minimal grain boundary angle $\Delta\theta_m$ 
is found to be independent of undercooling. The grain boundary energy 
for small angles is also almost independent of temperature (see 
Fig.~\ref{Tdependance} (b)). For high-angle grain boundaries, the grain 
boundary energy depends roughly linearly on the undercooling, which is 
expected since for such boundaries the center of the grain boundary is 
almost in the liquid state ($\phi\approx 0$), and the free energy
difference between solid and liquid is proportional to $T-T_m$.

We now turn to the case of negative undercooling (overheated solids). 
Where\-as it is difficult if not impossible to overheat solids in
experiments, due to the fundamental assymetry between solidification 
and melting (for a review, see \cite{Rettenmayr09}), the existence of
grain boundary solutions above the melting point has recently been
actively discussed in the context of grain-boundary premelting, due 
to new results coming from molecular dynamics \cite{Fensin10,Frolov11} 
and phase-field crystal \cite{Berry08,Mellenthin08} simulations. It
has also been examined in orientation-field \cite{Lobkovsky02,Tang06}
and multi-phase-field models \cite{Rappaz03,Mishin09,Wang10}. It seems
therefore  useful to analyze our model in the light of this
discussion.

A grain boundary can formally be described as two solid-liquid
interfaces separated by a thin disordered (almost liquid) layer. The
behaviour of a grain boundary when the melting point is approached
then depends on whether the grain boundary energy is larger or smaller
than the excess free energy of two solid-liquid interfaces. In the
former case, the grain boundary is {\it repulsive}, since at the
melting point, where the free energy density of both phases is equal,
it is advantageous to separate the two solids by a macroscopic
liquid layer. In the latter case, the grain boundary is {\it attractive},
since it is favorable for the two solids to stick together. More
quantitatively, the total free energy excess of the grain boundary
can be described as twice the solid-liquid surface free energy plus a
so-called {\it disjoining potential}, which describes the interaction
between the two interfaces as a function of their distance. In the
classic theory of grain boundary premelting, this disjoining potential
is assumed to be a simple exponential (see for example \cite{Rappaz03}),
but it was shown that this simple form is not sufficient to describe 
low-angle grain boundaries in the phase-field crystal 
model \cite{Mellenthin08}. Indeed, for these boundaries the interaction 
is attractive for large separations, but repulsive for short ones. As
a consequence, equilibrium grain boundaries of finite width can exist
at and above the melting point. It was later shown that such behavior
(long-range attraction and short-range repulsion) is also generic for
the standard multi-phase-field models (see \cite{Wang10} for a detailed
discussion).

In our model, the structure of the solution space can be directly and
simply deduced from the effective potential and the relation between
the constant $C$ and the total angle variation $\Delta\theta$. Let us
start out by a brief discussion of the solutions that exist when the
angle variation is zero. In this situation, $\partial_x\theta$ is zero
everywhere, which implies $C=0$, and the effective potential is 
just the negative of the double-well potential. 
For $T<T_m$, the solid has a lower free energy
than the liquid, which implies that the solid maximum in the effective
potential is higher than the liquid one. Therefore, no turning point
exists, and no grain boundary solution of the type outlined above is
possible. For $T>T_m$, the liquid maximum in $V_{\rm eff}$ is higher than the 
solid one, and therefore there always exists a turning point. The corresponding
grain boundary solution is unstable. Indeed, it can be seen as a thin
layer of liquid sandwiched between two solids, in a situation where
the attraction between the two solid-liquid interfaces is exactly
balanced by the free energy difference between solid and liquid.
Hence, this solution corresponds to a saddle point of the free energy,
which can be seen as a one-dimensional liquid nucleus. When the melting
point is approached from above, the thickness of the liquid layer diverges,
and the grain boundary energy approaches twice the solid-liquid surface
tension.

Next, consider the situation when $T<T_m$. If the angle variation is
finite, the constant $C$ differs from zero, and the effective potential
has two contributions. Since $g\to 0$ for $\phi\to 0$, the second term
in Eq. (\ref{veffdef}) diverges in this limit. In addition, 
since $g$ diverges for $\phi\to 1$,
$1/g$ vanishes for $\phi\to 1$. Since $C^2/g$ is positive, there is 
necessarily at least one solution for the equation 
$V_{\rm eff}(\phi_0)=V_{\rm eff}(1)$, which implies that there is a 
turning point for any value of $C$. It remains to link the values of
$C$ to the total angle variation. Whereas we did not find an analytical 
expression for this relationship, the main trends can be extracted from
an analysis of the trajectory close to the turning point. When $T$ tends
to the melting temperature, the constant $C$ necessarily becomes small 
since the liquid maximum in the returned double-well potential 
lies closely below zero.
This implies that the turning point is very close to the liquid state.
As a consequence, $g$ is also small, and the orientation gradient
becomes large. Since $g$ behaves like a power law with exponent larger
that one close to the liquid state (see appendix A), the gradient at
the center of the boundary becomes larger when $C$ becomes smaller.
Therefore, arbitrarily large total angle variations can be achieved
for any temperature below $T_m$.

This situation changes when $T>T_m$: now, the liquid maximum in the
returned double-well potential is higher than zero, 
which means that a turning point exists even for $C=0$.
When a non-zero value is chosen for $C$, this turning point moves 
further away from the liquid state. Therefore, the phase field remains
finite in the center of the grain boundary, and the orientation gradient
in the center of the grain boundary tends to zero when $C\to 0$.
In this case, the curve of $\Delta\theta$ versus $C$ is non-monotonous
and presents a maximum. This implies
that there is a maximum misorientation beyond which no grain boundary
solutions exist. This maximum misorientation depends on the temperature
and tends to infinity when $T\to T_m$. For misorientations smaller than
this critical value, there are two distinct solutions. The solutions 
for $C$ smaller than the maximum correspond to unstable liquid nuclei,
whereas the solution branch for $C$ larger than the maximum connects 
to the stable grain boundary solutions discussed previously.
Solutions corresponding to a grain boundary and an unstable liquid nucleus
are shown on Fig.~\ref{Tdependance} (d). While the angle profile is 
extremely similar in both cases, the liquid nucleus solution has a
much smaller minimal $\phi$. When the energies are plotted versus 
misorientation, the two branches meet at a cusp, as shown in 
Fig.~\ref{Tdependance} (b). At a fixed misorientation above the
minimal value discussed previously $\Delta\theta_m$, there are thus stable
grain boundary solutions for any $T<T_m$, and for a range of temperatures
$T>T_m$ that depends on the misorientation. This overheated range tends
to zero when the misorientation becomes large. 

The stable grain boundary solutions all exhibit a deviation of the phase field
from the solid state ($\phi=1$), which can be interpreted as a thin layer of
liquid by performing a Gibbs construction as in Ref.~\cite{Mellenthin08}.
The existence of such a layer of finite thickness results from the interplay
between two antagonistic effects, which can be understood through the interpretation
of the grain boundary as two solid-liquid interfaces that come close together.
On the one hand, the overlap of the two phase-field ``tails'' of the two interfaces
creates an attractive force that would eliminate the liquid layer (as is happens
in the case of zero misorientation). On the other hand, the finite total misorientation
generates a restoring force that opposes this compression of the liquid layer:
if the two interfaces get closer together, the orientation gradients have to
increase. On the branch on solutions that is stable, this increase of the gradient
leads to an increase of the free energy of the grain boundary solution, whereas
the opposite is true on the unstable branch.

As a result, all the grain boundaries in our model are attractive. This finding coincides 
with the classic criterion. Indeed, the solid-liquid surface tension can be calculated
analytically in our model and is equal to $\gamma_{\rm sl}=\sqrt{2}/6$. The curve of
$E_{GB}$ versus $\Delta\theta$ at $u=0$ ($T=T_m$), shown in
Fig.~\ref{Tdependance}, has an asymptote that tends exactly to 
$2\gamma_{\rm sl}=\sqrt{2}/3\approx 0.4714$ from below, which corresponds
to two solid-liquid interfaces with a weak attractive interaction that
is mediated by the exponential tails of the interfaces (see below). This 
behavior cannot easily be changed in our model. For example, the method used
in Ref.~\cite{Wang10} to make the interfaces repulsive is not applicable
here because it requires a free energy density with minima in both
field variables, which is not allowed if $\theta$ is to be interpreted 
as an orientation and rotational invariance is desired.

More complex behavior has been found in other orientation-field 
models \cite{Lobkovsky02,Tang06} by a suitable tuning of the various coupling
functions. In particular, for certain temperature ranges the possibility
of coexistence between two different grain boundary states was found.
Similar behavior could probably be generated in our model if the double-well
potential and the coupling function $g(\phi)$ are modified, but this
issue is not pursued any further here.

To summarize, the model presented here with an appropriate 
choice of the coupling function and double-well potential has stable 
localized grain-boundary solutions with an angle variation bigger 
than a threshold value (that can be made as small as desired by a 
proper choice of the quantity  $ab+d$). It also has solutions that 
correspond to a homogeneous solid with constant gradient in crystal 
orientation as long as this gradient is smaller than a threshold 
value. For a given total angle variation, the free energy of the 
grain boundary is always smaller than the energy of the homogeneous 
solution (which may be metastable). 

\begin{figure}
\includegraphics[width=0.4\textwidth]{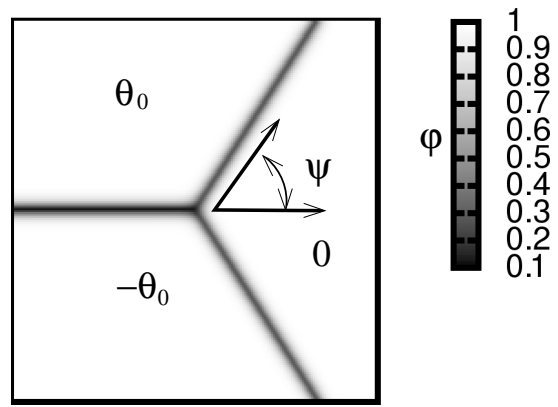}
\caption{\label{fig_trijunction}Grey scale plot of $\phi$ for an 
equilibrium trijunction. The three grains with respective grain 
orientation $\theta_0$, $-\theta_0$ and $0$ correspond to the white 
regions, and the grain boundaries correspond to the dark region.
Here, $\theta_0=10^\circ$, the undercooling was 0.085, and 
$\mu^2=11.2$.}
\end{figure}

\subsection{Trijunction points and Young's law}

We have also checked explicitly that Young's law is satisfied
at trijunction points. In order to obtain a static trijunction
and to avoid curvature effects induced by simple Neumann 
boundary conditions, the simulation was performed as follows: 
first, we let three initially circular grains grow in a simulation 
box with Neumann boundary conditions. When the grains have impinged
on the system boundaries, we use the value of the fields at the boundary
at a certain time as Dirichlet boundary condition for the further 
evolution, which imposes that neither the phase field 
nor $\theta$ evolves any further at the boundary. 
This amounts to pinning the grain boundaries at fixed 
points of the system boundary. For initial orientations of the three
grains given by $\theta_0$, $0$, and $-\theta_0$ as displayed in
Fig.~\ref{fig_trijunction}, Young's law predicts
\begin{equation}
\psi=\mathrm{acos}\left(\frac{E_{GB}(2\theta_0)}{2E_{GB}(\theta_0)}\right),
\end{equation}
where $\psi$ is the half of the dihedral angle, as sketched in
Fig.~\ref{fig_trijunction}; $E_{GB}(\theta_0)$ and $E_{GB}(2\theta_0)$
are the energies of grain boundaries with misorientations $\theta_0$
and $2\theta_0$ that have been calculated in the previous subsection.
For $\theta_0=4.2^\circ$ (resp. $10^\circ$), the angle $\psi$ 
was $46^\circ$ (resp. $55^\circ$), to be compared to $46^\circ$ (resp.
$52.4^\circ$) predicted from Young's law. The agreement is excellent,
which is not surprising because Young's law should generally be
valid in phase-field models that derive from an energy 
minimization \cite{Garcke98}.

\begin{figure}
\begin{center}
	\includegraphics[width=0.5\textwidth]{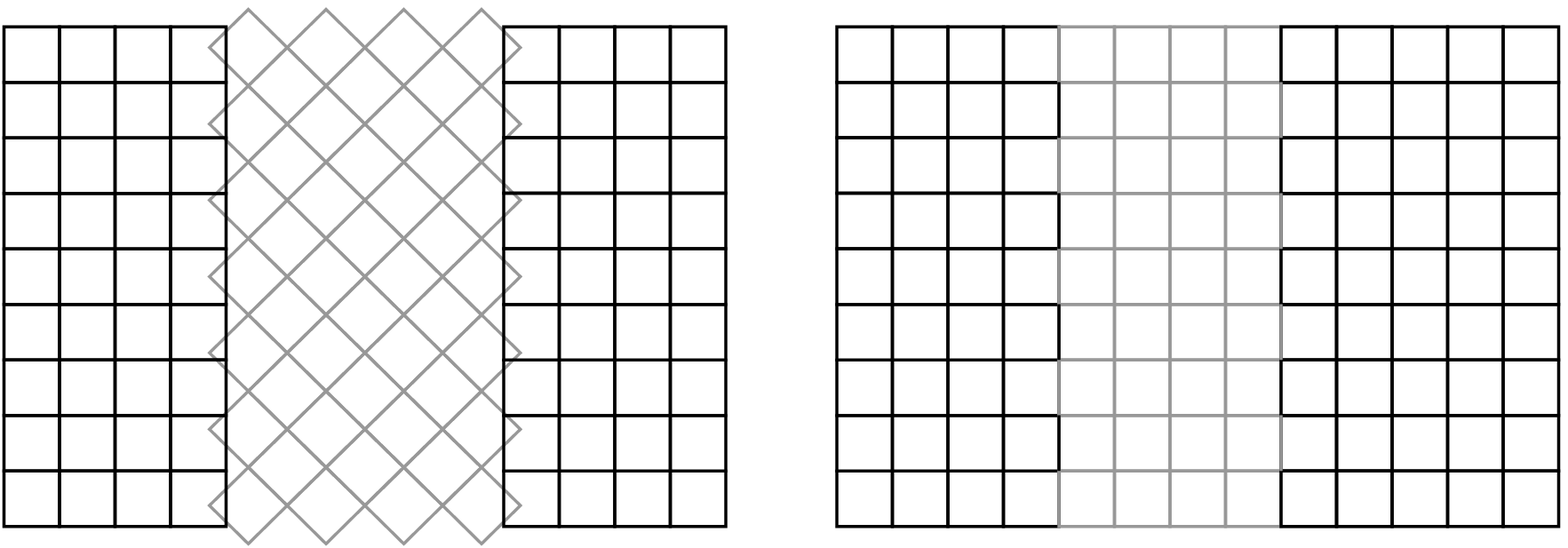}\\
	\includegraphics[width=0.5\textwidth]{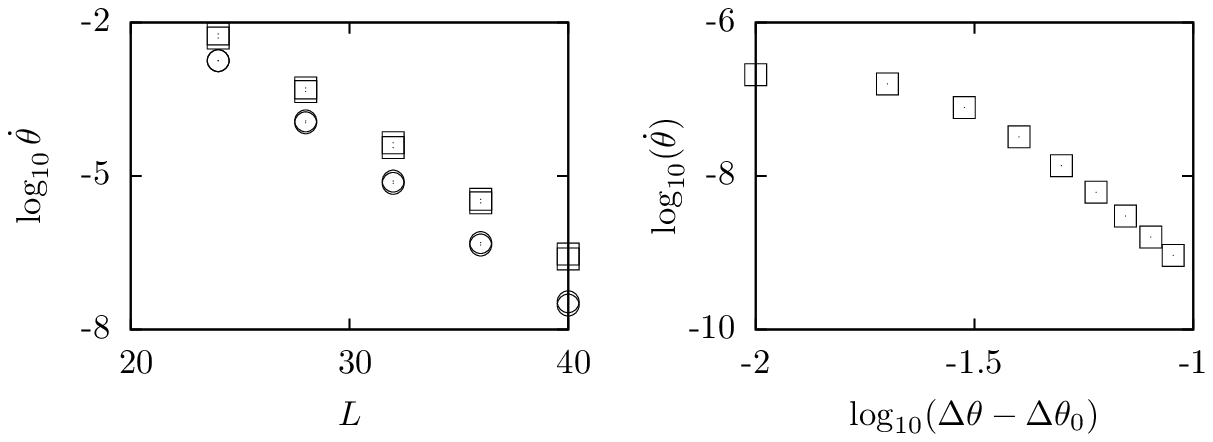}
\end{center}
\caption{\label{fig_tricristal}\textbf{Top:} Sketch of the tricristal
configuration, which can evolve towards a homogeneous system by the
rotation of the central slab. \textbf{Bottom:} Rotation rate of the
central slab as a function of distance between the boundaries  for two different
values of the angle difference through the grain boundary ($\Delta \theta$) (left)
and as a function of misorientation ($\Delta \theta$) for a given value of the
distance between the two grain boundaries (right). In these simulations, the
orientation of the outer grains is kept constant.}

\end{figure}

\subsection{Grain boundary tails and the tricristal}

Next, we briefly examine the interaction of two grain boundaries in 
the {\em tricristal} configuration sketched in Fig.~\ref{fig_tricristal}:
two grains of the same orientation sandwich a slab of material with a
different orientation. The two grain boundaries obviously contribute
a finite free energy. Since the evolution of the system is variational
according to Eqs. (\ref{phi_evolution}) and (\ref{theta_evolution}) and nothing 
in the equations forbids the central crystal slab to rotate, the system tends
to eliminate these defects: the orientation of the central slab slowly 
changes with time to approach the one of the outer grains, until the
angle field finally becomes homogeneous. Obviously, for the sketched
geometry this change cannot correspond to rigid body rotation. The
evolution equation for the orientation actually describes the {\it local}
rotation of disconnected objects, such as the molecules of a liquid
crystal. For a continuously connected rigid solid, such a rotation
could only be accomplished by a change in local connectivity,
obtained for example by the migration of dislocations from one grain
boundary to the other (see \cite{Plapp11} for a more detailed 
discussion). However, since this process (if it takes place at all) 
would be extremely slow, the rotation of the central slab
should be suppressed. In the previous orientation-field models 
that contain a modulus of the gradient in the free energy functional, 
this is problematic since the functional derivative has a singularity 
at the points with vanishing orientation gradients -- that is, 
inside the grains. The singular diffusion equation that results 
from the functional derivative leads to a non-local coupling 
of the grain boundaries \cite{Kobayashi99}. As a consequence, 
the rate of rotation of the central crystal slab does not depend on
the distance between the grain boundaries. Therefore, the only 
solution to suppress the unwanted rotation is to choose a 
mobility function which vanishes inside the crystal grains \cite{Warren03}.

In our model, the interaction between the grain boundaries is local.
This can be easily demonstrated by an analysis of the ``tails'' of the
boundary, that is, the approach of the phase and orientation fields
to their bulk values. This is again straightforward by using the
mechanical analog. Indeed, a linearization of Eq. (\ref{eq_phi_eff})
in the vicinity of the solid state yields
\begin{equation}
\phi_{xx}=-\left.\frac{\partial^2 V_{\rm eff}}{\partial \phi^2}\right|_{\phi=1}(\phi-1),
\end{equation}
the solution of which is
\begin{equation}
\phi(x)=1-A\exp[\pm\xi(x-x_0)]
\end{equation}
with $\xi=\sqrt{\left.(\partial^2 V_{\rm eff})/(\partial \phi^2)\right|_{\phi=1}}$, 
and the values of the integration constants $A$ and $x_0$ as well as the 
sign inside the 
exponential are fixed by the boundary conditions (which depend on the
position of the grain boundary). The gradient in orientation is then obtained
from Eq. (\ref{eq_thetaprime}) and tends to zero in the grains since 
$g(\phi)$ diverges as $\phi\to 1$.

Several conclusions can be drawn from this result. First, the grain boundaries
in our model are truly localized, in the sense that in a heterogeneous system
only an exponentially small part of the orientation variation takes place
outside the grain boundaries. Second, the interaction between two grain
boundaries decays exponentially with their distance for well-separated grain
boundaries, since for large separation the interaction is mediated by the overlap
of the exponential tails (see \cite{Wang10} for a more detailed discussion).
As a consequence, even for a constant mobility in the orientation evolution
equation ($Q(\phi,\nabla\theta)=1$), the rotation rate of the central slab in the tricristal configuration
should rapidly drop with the size of the slab. Finally, the characteristic length
of decay for the exponential tails $1/\xi$ is the inverse of the second derivative
of the effective potential, taken in the solid state. As discussed above, the
second derivative depends of the value of the constant $C$ and vanishes at a
critical value of $C$ that corresponds to the minimal misorientation $\Delta\theta_m$.
Therefore, grain boundaries with smaller misorientations interact more strongly.

These predictions are borne out by simulations on the tricristal
configuration, as shown in Fig.~\ref{fig_tricristal}. For grain boundaries 
of a fixed misorientation, the rotation rate of the central slab decays
exponentially with systems size (and thus, separation of the grain boundaries),
and the rotation rate is larger for smaller misorientations at a fixed
separation. Therefore, while our model does permit grain rotation,
qualitatively by the same mechanism than in previous orientation-field models,
the artefacts introduced by this design are small even for a constant
mobility function, and it can be hoped that they will not drastically 
affect the results of large-scale simulations.

\section{Anisotropic interfaces}
\label{sec_anis}
Our model can describe two different types of boundaries:
solid-liquid interfaces and grain boundaries, both of which
are in general anisotropic. For solid-liquid interfaces, the surface
tension and interface mobility depend on the orientation of the
interface with respect to the crystal axes of the solid. For grain
boundaries, the energy and mobility depend on the orientation of
{\em both} crystals that meet at the boundary. This yields two 
independent parameters in two dimensions, and five independent 
parameters in three dimensions. Grain boundary energies
and mobilities are conventionally given as functions of the 
{\em misorientation} between the two crystals and the {\em inclination} 
of the grain boundaries. 

The standard way of including the anisotropy of the solid-liquid interfaces
in phase-field models is to make the coefficient of the phase-field gradient 
term dependent on the interface orientation (see for example \cite{KarRap98}). 
We will now explore the effects of introducing this type of anisotropy in our 
model. As we will see below, the modification of the phase-field 
gradient term also makes grain boundaries anisotropic. Since the
anisotropies in both types of boundaries are created by one
single term in the free energy functional, they are obviously not
independent. A possible way to choose both anisotropies independently 
would be to introduce a suitable dependence on orientation in the 
coefficient of the orientation-field gradient term. However, this 
possibility will not be explored further here.

For a crystal of a cubic material in two dimensions with its 
crystallographic axes aligned with the coordinate system, the 
orientation-dependent surface tension is usually written as
\begin{equation}
\gamma(\varphi) = \gamma_0\left[1+\epsilon_4\cos(4\varphi)\right],
\end{equation}
where $\varphi$ is the angle between the interface normal (pointing into
the liquid) and the $x$ axis. If the crystal is rotated by an angle $\theta$,
this expression becomes
\begin{equation}
\gamma(\varphi,\theta) = \gamma_0\left[1+\epsilon_4\cos[4(\varphi-\theta)]\right].
\end{equation}
To implement this anisotropy in the phase-field model, we define the
unit normal vector by $\mathbf{n}=-\nabla\phi/|\nabla\phi|$ and choose
the gradient coefficient to be
\begin{equation}
W_\phi(\mathbf{n},\theta) = W_0 a(\mathbf{n},\theta)
\label{wchoice}
\end{equation}
with the anisotropy function
\begin{eqnarray}
a(\mathbf{n},\theta) & = & 1+\epsilon_4\left[
          \cos(4\theta)\left(4(n_x^4+n_y^4)-3\right)\right. \nonumber\\
  & & \qquad \mbox{} +
          \left.\sin(4\theta)\left(4n_xn_y^3-4n_yn_x^3\right)\right],
\end{eqnarray}
where we have repeatedly used the addition theorems for the 
trigonometric functions. Lengths are now scaled by the average interface
thickness $W_0$. 

A few consequences of the fourfold symmetry should be mentioned here.
The fundamental issue is that for a cubic material all orientations
that differ by integer multiples of $\pi/2$ are equivalent. Indeed,
a crystal unit cell can be locally rotated by a multiple of $\pi/2$
without changing the state of the crystal. Therefore, $\theta$ is 
actually defined modulo $\pi/2$, which means that the values of
$\theta$ can be restricted to the interval $[0,\pi/2[$. This means
that, contrary to what we did in the previous section, we cannot 
rescale $\theta$ to make $\mu=1$. Nevertheless, we will consider
in the following the case $\mu=1$ for simplicity. Furthermore,
care has to be taken in the calculation of differential operators.
For instance, in the evaluation of the gradient energy and in
the Laplace operator, differences between the values of $\theta$
at neighboring grid points need to be computed. We follow the
method proposed in Ref.~\cite{Pusztai2005}: in the evaluation 
of such operators for a given grid point $i$, we check for each
neighboring point which of all the equivalent values of 
$\theta$ gives the lowest value of the gradient energy, and 
calculate the evolution of point $i$ using this value. This
procedure ensures that local ``jumps'' of the angle field with
amplitude $\pi/2$ (or multiples of it) do not introduce any
energy cost. Thus, all the crystal symmetries are correctly
implemented. Nevertheless, for simplicity of exposition, we will 
use the standard notation for differential operators in the 
remainder of the paper.

The evaluation of the variational derivatives in Eqs. (\ref{phi_evolution}) 
and (\ref{theta_evolution}) yields
\begin{eqnarray}
P(\phi,\nabla\theta)\partial_t\phi & = & \nabla\cdot \left[a(\mathbf{n},\theta)^2\nabla\phi\right] \nonumber \\
 & & \mbox{} +\partial_x \left(|\nabla\phi|^2a(\mathbf{n},\theta)\frac{\partial a(\mathbf{n},\theta)}{\partial(\partial_x\phi)}\right) \nonumber \\
 & & \mbox{} +\partial_y \left(|\nabla\phi|^2a(\mathbf{n},\theta)\frac{\partial a(\mathbf{n},\theta)}{\partial(\partial_y\phi)}\right)  \nonumber \\
 & & \mbox{} - V'(\phi,T) - \mu^2(\nabla\theta)^2g'(\phi)
\label{anis_phi}
\end{eqnarray}
and
\begin{equation}
Q(\phi,\nabla\theta)\partial_t\theta =
 \mu^2\nabla\cdot\left(g(\phi)\nabla\theta\right)
  - a(\mathbf{n},\theta)\frac{\partial a(\mathbf{n},\theta)}{\partial\theta} (\nabla\phi)^2.
\label{anis_theta}
\end{equation}
For a uniform (constant) orientation field, the first of these equations 
is identical to the standard anisotropic phase-field equation 
of Ref.~\cite{KarRap98}, with the crystalline axes rotated by
the angle $\theta$ away from the $x$ axis. The equation for
the orientation field contains the derivative with respect to the
angle of the phase-field gradient coefficient. Physically, this
term represents a ``torque'' exerted on the orientation field
at the surface. Indeed, if the solid-liquid interface is anisotropic,
this term drives the angle field towards an orientation which
corresponds to a minimum of the surface free energy.

In the following, we will examine in more detail the influence 
of the new terms on grain boundaries, solid-liquid interfaces,
isolated monocrystals and polycrystals. 

\subsection{Grain boundaries}
\begin{figure}
\centerline{
\includegraphics[width=0.5\textwidth]{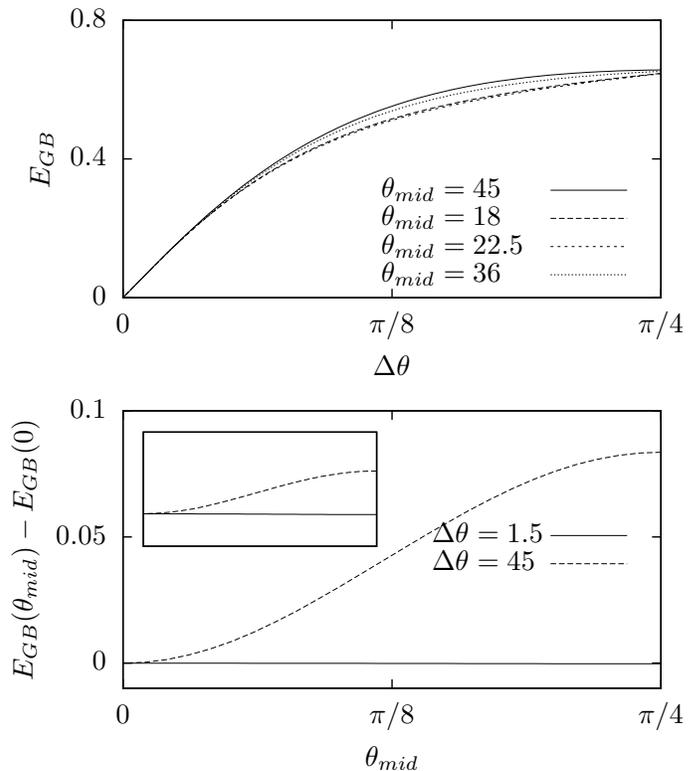}}
\caption{\label{fig_anisotropie}\textbf{Top:} Grain boundary energy 
as a function of the misorientation for different values of the median 
angle ($\epsilon=0.05$, $u=-0.1$. \textbf{Bottom}: grain
boundary energy as a function of the median angle for two different 
values of the misorientation. Inset: same curves, normalized by the 
minimal energy of such a GB ($\approx 0.5$ for the case where the 
misorientation is $1.5^\circ$ and $\approx 3.2$ for a $45^\circ$ 
misorientation). Clearly, for small values of the misorientation, 
the grain boundary energy is less sensitive to anisotropy than for 
high values.}
\end{figure}

We have systematically calculated stationary one-dimensional grain 
boundary solutions of the anisotropic equations (\ref{anis_phi}) and
(\ref{anis_theta}), and determined the grain boundary energy. 
The results are presented in Fig.~\ref{fig_anisotropie}. As
stated above, in two dimensions, there are two independent parameters.
For two grains of respective orientations $\theta_1$ and $\theta_2$,
we define the misorientation, as before, as $\Delta\theta=\theta_2-\theta_1$,
and the median angle as $\theta_{\rm mid} = (\theta_1+\theta_2)/2$.

The curve of the grain boundary energy versus the median orientation 
for a fixed misorientation exhibits the expected behaviour, that is, a
$4\theta$ dependence shown by the $\pi/2$ periodicity of the grain boundary
energy as a function of $\theta_{\rm mid}$. We limit ourselves to inclinations 
that are smaller than $\pi/2$ since other values can be reached by symmetry.
It is also interesting to note that higher misorientations can have a
lower grain boundary energy for certain inclinations. This is due 
to the fact that a grain boundary consists, roughly speaking, of two
solid/liquid interfaces and that changing the inclination 
rotates both of these interfaces with respect to the orientation of
the crystalline axes. This can lead to decrease of the energetic cost 
that can be higher than the increase due to an increase of the
misorientation.

It should be mentioned that in our first implementation of the
model, we noticed a very slow motion of one-dimensional anisotropic 
grain boundaries \cite{Mellenthin07}. This is clearly unphysical
since the free energy density in the grains on both sides is identical, 
which precludes a persistent motion to one side. We found that this 
effect was due to the discretization. 
Indeed, the anisotropic terms lead to assymetric $\phi$ profiles at the grain
boundary since the grain orientation relative to the grain boundary normal is,
in general, not the same on both sides of the interface. Therefore, 
the unavoidable discretization errors present on both sides of the 
boundary do in general not exactly compensate each other, which then 
leads to an unwanted movement of the interface.
We eliminated this problem by changing the discretization scheme:
instead of discretizing the equations of motion (\ref{anis_phi})
and (\ref{anis_theta}), we discretized the free energy functional and
replaced the functional derivatives by ordinary derivatives. As
a result, we obtain discrete equations that derive from a discrete
energy function. It can easily be shown (see appendix B) that with 
this scheme the energy function is strictly decreasing with time;
indeed, no spurious motion of the grain boundaries took place with
this modified scheme. 

\subsection{Solid-liquid interfaces and equilibrium crystal shapes}
Next, we turn to the study of solid-liquid interfaces. Bulk liquid
and solid coexist at the melting point ($T=T_m$, $u=0$), which implies
that there must exist an equilibrium solution for a planar interface.
For a constant orientation field ($\theta(\mathbf{x})=\theta_0$),
this solution (for an interface normal to the $x$ direction and
centered at position $x_0$, with the solid located in the domain
$x<x_0$) can be obtained analytically and reads
\begin{equation}
\phi(x) = \frac 12 
  \left[1 - \tanh\left(\frac{x-x_0}{\sqrt{2}W_\phi(\theta_0)}\right)\right].
\label{intsolution}
\end{equation}
This interface has a surface free energy of 
$\gamma=\gamma_0W_\phi(\theta_0)/W_0$, which has already been used to
choose $W_\phi$ in Eq. (\ref{wchoice}).

However, a constant orientation field is a stationary solution of
the anisotropic orientation equation (\ref{anis_theta}) only if the
angle $\theta_0$ corresponds to symmetry direction of the crystal
(a minimum or maximum of the anisotropy function). In all other 
cases, the ``torque'' term is non-zero inside the interface, which
generates an evolution of the orientation if the system is
started from a constant orientation field.

This effect is actually not physical for a crystalline material. It is 
completely analogous to the {\em anchoring} effect known for liquid 
crystals. Indeed, in such materials, the constituent molecules are 
anisotropic and exhibit an orientational order, but are free to rotate
individually. Therefore, if the surface energy depends on the orientation 
of the molecules with respect to the surface normal, a competition between 
bulk and surface effect takes place. Any variation of the orientation in the 
bulk is energetically penalized, but the energy gain due to the surface 
terms can offset this penalty for molecules close to the surface, 
such that the orientation tends to an energetically favorable 
direction close to the surface. In contrast, in a crystalline material,
the structure cannot change by rotation of the individual crystalline unit cells. 
The interface orientation can change only by attachment and detachment 
of atoms.

To eliminate this effect, existing orientation-field models adopt a radical 
strategy \cite{Granasy04}: they just set the ``torque'' term to zero and 
use Eq. (\ref{anis_theta}) with only its first term on the right hand side. 
Whereas this procedure can be justified by the arguments given above, 
it breaks the variational structure of the model, which means that several
desirable mathematical properties are lost (in particular, the free energy
is not guaranteed to decrease with time any more). Instead of adopting
the same strategy in our model, we will proceed by showing that the effects 
induced by this surface term are actually small. Once again, 
the facts that the diffusion part of the orientation equation is {\em regular} 
and that the orientation variation is localized in the vicinity of the 
interfaces are crucial for this property of the model.

When we start the two coupled equations (\ref{anis_phi}) and
(\ref{anis_theta}) from the solution given by Eq.~(\ref{intsolution})
at a fixed temperature $u=0$, we observe that the solid grows
very slowly at the expense of the liquid. This can be qualitatively
understood by the following reasoning. As mentioned above, the
torque term induces an evolution of the orientation field towards
a more favorable direction. If the solid could ``turn'' by changing
its orientation, the system would be able to reach its global
energy minimum (the orientation field aligned with a direction
of minimal surface tension). However, since the gradient of
orientation tends exponentially to zero away from the interfaces,
there is actually no term in the orientation equation which could
induce such a rotation. However, the system can lower its energy by 
moving the interface and simultaneously changing the orientation
field inside the interface. In this way, a homogeneous liquid is successively
replaced by a solid with a very small ``frozen in'' orientation gradient
that results from the time evolution of the surface orientation. 
This ``strained'' solid has a higher free energy density than the liquid, 
but this can be offset by the lowering of the surface free energy, 
such that the total energy of the system still decreases.

\begin{figure}
\centerline{
\includegraphics[width=0.5\textwidth]{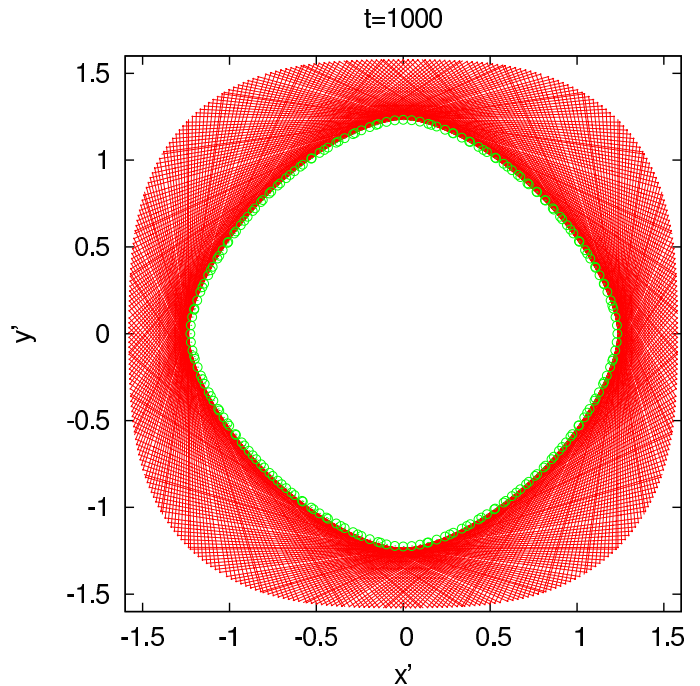}}
\caption{\label{fig_wulff}(Color online) The equilibrium shape of an anisotropic
crystal (green circles) agrees well with the Wulff shape (inner envelope
of the red lines). The anisotropy is $\epsilon_4=0.05$.}
\end{figure}

In summary, the anisotropy of the solid-liquid interface induces
an unphysical growth of the solid at the melting temperature. In other 
words, the presence of the additional degree of freedom (change in 
orientation) leads to a shift of the melting point for anisotropic
interfaces. To quantify the importance of this effect, we have
determined the modified melting temperature by performing
simulations at constant volume of solid, as described in appendix C.
The shift is actually very small, of the order of $u\sim 10^{-4}$.
Moreover, since the profile of phase field and orientation across
the interface are not exactly given by Eq. (\ref{intsolution}), the
actual surface tension slightly differs from the intended value.
To test the magnitude of this effect, we have determined the
equilibrium shape of a two-dimensional anisotropic crystal, again
by performing simulations at constant volume of solid. In 
Fig.~\ref{fig_wulff}, we show the resulting equilibrium shape and
the comparison to the Wulff construction, which is very satisfactory.
Indeed, for typical undercoolings that can be reached in simulations
($u\sim 0.1$), the shift of the equilibrium point is negligibly
small, and therefore it is not surprising that the model is capable
of producing an excellent approximation to the equilibrium crystal shape.

We have not explicitly checked Young's law at trijunction points
for the anisotropic version of our model. However, it is shown 
in Ref.~\cite{Garcke98} that Young's law is generally valid for
phase-field models that are variational, since equilibrium states
result from an energy minimization. Since this is the case for
our model, and the equilibrium crystal shape is well reproduced,
we are fairly confident that our model also correctly implements
the balance of Herring torque terms at trijunction points.

\subsection{Polycrystals}
\begin{figure}
\centerline{
\includegraphics[width=0.5\textwidth]{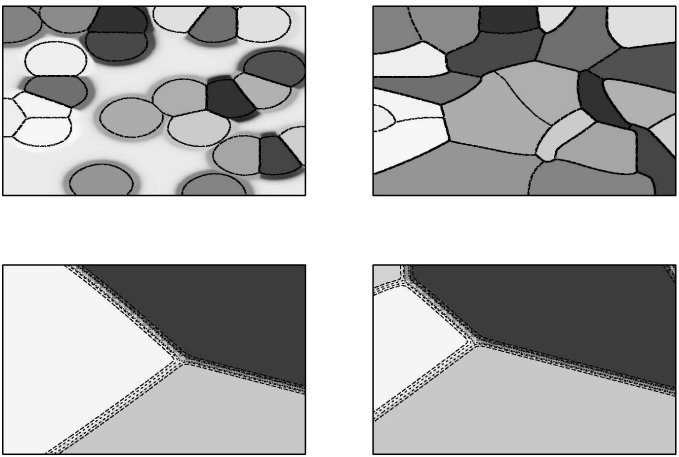}}
\caption{ \label{fig_murissement}
Snapshots of the numerical simulation of the isothermal growth of multiple
grains. The grey scale indicates the orientation of grains.
The undercooling is kept homogeneous and constant and the initial grains
(top left) grow until they begin to interact with each other (top right).
Once all the liquid is transformed, the further evolution is only driven 
by differences in grain boundary energies and becomes much slower. 
As one can see, most grain boundaries are already straight (especially
high angle grain boundaries between dark and light regions). A typical 
change in structure is shown in the two plots at the bottom: snapshots 
at two successive times of the same region (only a small part of the 
system is shown) are displayed. Between the two snapshots, the length 
of the high angle grain boundary (between the dark and light region) 
has significantly decreased, which has led to a motion from right to 
left of the triple junction.}
\end{figure}

We now turn to a more illustrative part, where we show results 
of numerical simulations for the case of isothermal solidification
(that is, we set $u$ to a constant that does not evolve with time). 
The aim is to show that our model is capable of reproducing the 
evolution of polycrystals, at least qualitatively. Hence, we have 
simulated the solidification of an undercooled melt with a few seeds 
of different crystalline orientations. The result is shown in 
Fig. \ref{fig_murissement}. First, the individual grains grow. Since
we do not solve the heat equation (\ref{uevolution}), no diffusive 
instability can
appear and create dendrites, and the shape of the grains remains
convex. Once the grains impinge, grain boundaries appear. They evolve 
toward straight segments, the lengths of which change through
the motion of trijunctions while the orientation in each grain 
remains unchanged. Other situations such as directional solidification 
in a fixed temperature gradient were also simulated, and also
led to qualitatively correct behaviour of the model.
   
\section{Conclusion}
\label{sec_end}
We have presented a phase-field model for the solidification and coarsening of
polycrystals that is formulated in terms of two continuous fields, a phase field
that indicates the local state of matter and an orientation field that gives the
local direction of the crystallographic axes. Contrary to previous orientation-field
models found in the literature, the free-energy functional of our model does not
contain a term proportional to the modulus of the orientation gradient, but only
a standard square gradient term. Stable localized grain boundary solutions 
are instead generated by a singular coupling function between the orientation 
and the phase  field.

As a result, the mathematical structure of the model and the evolution equations
are closer to the standard phase-field models for solidification, in that the evolution
equation for the phase field is a regular reaction-diffusion equation instead of a
singular diffusion equation that has to be regularized in a suitable way. The properties
of the grain boundary solutions can be investigated by standard mathematical
methods, in particular by the use of the well-known mechanical analog that exploits
the relation between interface solutions and the motion of a particle in a potential 
landscape. This method has allowed us to perform a thorough analysis of the
grain boundary properties, and to choose a suitable mathematical form of the
singular coupling function.

One important result of this analysis is that the grain boundary solutions of our model
have a similar structure as the solid-liquid interfaces in standard phase-field models: 
a central region with strong gradients of both fields is surrounded
by exponential tails, and the interaction between two grain boundaries becomes
exponentially small with their distance. Two important differences with previous
orientation-field models are a consequence of these facts. First, the artificial
grain rotation that is mediated by long-range interactions between grain boundaries
is not present in our model. Second, it is possible to incorporate interfacial
anisotropy in our model without breaking the variational structure of the model.
The anisotropy generates unphysical ``torque'' terms that drives the orientation
fields towards the minima of the surface free energy, but these effects are
negligibly small for typical simulation parameters that can be achieved in practice.

We have presented some preliminary simulations performed in two dimensions, 
which show that a qualitative description of multi-grain growth and of the evolution 
of polycrystalline solids can be obtained using this model. It should be
stressed that the evolution equation for the orientation obtained from the
variational formalism, Eq. (\ref{theta_evolution}), is likely to be incorrect, as 
discussed in more detail in Ref. \cite{Plapp11}. Therefore, it is unlikely that this
model can be used for quantitative simulations of grain growth. Nevertheless,
in our opinion this model is interesting because of its intrinsic mathematical
structure and because of its ability to reproduce a host of complex pattern-formation
phenomena with the help of a simple set of equations.

We have concentrated here on the discussion of the mathematical structure of
the model, but we have neither tested its efficiency nor its quantitative
accuracy in the case of known model problems, such as the growth of an
isolated dendrite \cite{KarRap98}. These are promising subjects for further studies.

\appendix

\section{Choice of model parameters: further details \label{furtherdetails}}
The choice of the coupling function for values of the phase field close 
to one (the solid) has already been discussed in the main text. 
Here, we briefly discuss the form of the coupling function $g$ close 
to zero and the double-well potential used in our simulations. First, 
$g$ must be equal to zero in the liquid (there should be no energy 
penalty associated with orientation changes in the liquid phase). Furthermore, 
we must have $g'(0)=0$ in order to keep a minimum of the free energy at
the liquid state, irrespective of the configuration of the orientation
field, which has no meaning inside the liquid. Finally, we also require
that the variation of the orientation should remain confined in the
grain boundary. A similar problem was solved in a phase-field model of
fracture, where the strain must be localized inside the crack for a broken
elastic material. According to the results of Ref. \cite{Karmafrac}, this
imposes that $g$ should behave as $\phi^\beta$ for $\phi\to 0$, with 
$\beta>2$. Choosing $\beta=2$ would lead to a singular behaviour of $\theta$ 
in the middle of the grain boundary (and to strong pinning of the grain
boundary on the grid points). Therefore, we choose $\beta=3$.

The conditions on the double-well potential are the following:
it should have two minima at $\phi=0$ and $\phi=1$ for all temperatures,
with a free energy difference proportional to $u$ between the two wells.
Furthermore, its second and third derivatives in the solid should be
constant; this is guaranteed if its leading order expansion in $\phi=1$ is of 
the form $(1-\phi)^2-2(1-\phi)^3$; with this choice,  Eq.~\ref{eq_1dhom_ineq_dev}
is satisfied. In addition, we want to prescribe both the second derivative of the 
potential $C_{\rm liq}$ and its value $V_0$ in $\phi=0$. The first is set to the
same value as the second derivative in the solid well, in order to guarantee a
symmetric interface profile of the phase field, and the second is a function of
temperature, whereas the depth of the solid well is independent of temperature. 
Assuming that $V$ has a polynomial form, the number of equation it has to satisfy is 7:
\begin{enumerate}
\item{The free energy of the liquid phase is $V_0$ : $V(0,T)=V_0(T)$}
\item{The liquid phase is a minimum	: $V'(0)=0$}\
\item{The curvature of the potential at $\phi=0$ is fixed :
	$V''(0)=2C_{\mathrm{liq}}$}
\item{The free energy of the solid state is 0:$ V(1)=0$}
\item{$\phi=1$ is a minimum : $V'(1)=0$	}
\item{The  expansion of $V$ in $\phi=1$ has to be  $(1-\phi)^2-2(1-\phi)^3$ 
	: $V''(1)=2$ \underline{and} $V^{(3)}=12$}
	\end{enumerate}
The first 3 equations impose the value of the constant and of the coefficients 
$\phi$ and $\phi^2$ in the polynomial expansion of $V(\phi)$, while the last four
equations (the last three points of the above enumeration) can be easily
translated into a set of four independent linear equations on the coefficient of
$V$. An evident solution is then to consider that $V$ is a $6^{th}$ order
polynomial and to solve the linear system imposed by the last four equations in
order to compute the remaining coefficients of $V$. 
This approach leads to the following
expression  of V:
\begin{equation}
V(x)=V_0+C_{\rm liq}x^2+cx^3+dx^4+ex^5+fx^6.
\end{equation}
The remaining 4 coefficients have then to satisfy a set of 4 independent linear
equations that can be inverted giving the following 
polynomial form for $V$~:
\begin{widetext}
\begin{equation}
\begin{array}{l}
V(x)=U_{liq}+C_{\rm liq}x^2+cx^3+dx^4+ex^5+fx^6.\\
\left(
\begin{array}{c}
c\\
d\\
e\\
f
\end{array}
\right)
=\frac{1}{6}\left(	

\begin{array}{cccc} 
 	
 120  & 	
 -60  &  	
12  &  	
-1   \\
 	
 -270  & 	
 150  & 	
 -33  &  	
3   \\
 	
 216  & 	
-126  & 	
30   &  	
-3  \\
  	
-60  & 	
36  &  	
-9 & 
2/3   
\end{array}
\right)
\left(
\begin{array}{c}
	
-V_0-C_{\rm liq}\\
 
-2C_{\rm liq}\\

2-2C_{\rm liq}\\

12
\end{array}
\right)
\begin{array}{c}
(V(1)=0)\\
(V'(1)=0)\\
(V''(1)=2)\\
(V^{(3)}(1)=12)
\end{array}\label{coef_V}
\end{array}
\end{equation}
 Choosing to have $C_{\rm liq}=1$, we have:
\begin{equation}
V(\phi)=V_0+\phi^2-(2+20V_0)\phi^3+(1+45V_0)\phi^4-36V_0\phi^5+10V_0\phi^6
\label{def_V_T}
\end{equation}
\end{widetext}
The choice of $V_0=-\lambda u$ (which yields the correct free energy difference between
the two bulk phases) gives Eq. (\ref{vfunc}). With this choice of $V$, a reasonable choice for 
$g$ is finally:
\begin{equation}
g=\frac{7\phi^3-6\phi^4}{(1-\phi)^2}
\end{equation}

\section{Discretization issues}
Here, we illustrate briefly how a discretization scheme that does not derive
from an energy can lead to spurious motion of the interface while a
discretization scheme that derives from a free energy cannot lead to any
such effect. The free energy we consider (for simplicity, in one dimension) writes
\begin{equation}
	 \mathcal{F}=\int dx \varepsilon(\theta) \frac{(\nabla
	 \phi)^2}{2}+V(\phi)+g(\phi)(\nabla \theta)^2
\end{equation}
When simulating the behaviour of a partial differential equation (PDE) that derives from 
this free energy, one can either first derive the PDE and then discretize it or first discretize 
the free energy and then derive the evolution equation for the discretized field.
In the following, we consider both options and show how they differ in the case
of the first term in the above functional. The evolution equation that derives from this 
term for the phase field $\phi$ is
\begin{eqnarray}
	\partial_t \phi&=&\partial_x (\varepsilon(\theta)\partial_x
	\phi)\label{PDE1}\\
	&=& \varepsilon'(\theta)\partial_x \theta \partial_x
	\phi+\varepsilon(\theta)\partial_x^2 \phi \label{PDE2},
\end{eqnarray}
where we have used the chain rule to obtain the second expression.
The latter equation is simply discretized as follows:
\begin{eqnarray}
	\partial_t{ \phi_n}&=&\nonumber
	\varepsilon'(\theta_n)\frac{(\theta_{n+1}-\theta_{n-1})(\phi_{n+1}-\phi_{n-1})}{4(\Delta x)^2}
\label{disc_from_PDE_div}\\
	&+&
	\varepsilon(\theta_n)\frac{\phi_{n+1}-2\phi_n+\phi_{n-1}}{(\Delta x)^2}
	\label{disc_from_PDE}\end{eqnarray}
  The other approach is to first discretize the free energy:
	\begin{equation}
		\mathcal{F}_d=\sum_n { \varepsilon(\theta_{n+1/2}) \over 2 }
		\left(\frac{\phi_{n+1}-\phi_{n}}{\Delta x}\right)^2, 
	\end{equation} 
where $\theta_{n+1/2}$ is either the value of $\theta$ on a staggered grid or
the mean value of $\theta$ between the grid points.
This, using the relaxation law $d{\phi_n}/dt=-\partial\mathcal{F}_d /
\partial\phi_n$, leads  to an evolution equation for $\phi_n$ that writes
(taking into account all the terms of the sum):
\begin{equation}
		\frac{d\phi_n}{dt}=\varepsilon(\theta_{n+1/2})\frac{\phi_{n+1}-\phi_n}{(\Delta x)^2}+\varepsilon(\theta_{n-1/2})\frac{\phi_{n-1}-\phi_n}{(\Delta x)^2}
		\label{disc_from_energy}\end{equation}
This equation differs from Eq.~(\ref{disc_from_PDE}) by little. Indeed, it is
easy to see that it corresponds to the discretization of Eq.~(\ref{PDE1}) 
instead of Eq.~(\ref{disc_from_PDE_div}). Using a Taylor expansion centered at 
the grid point $n$ one can show that both discretization schemes are equivalent up 
to the second order, but differ by terms of higher order. This just reflects the fact
that the chain rule is valid for continuous fields, but not for discrete fields. 

These seemingly small differences have important consequences. Due to 
its construction, Eq.~(\ref{disc_from_energy}) implies that 
$\mathcal{F}_d$ is decreasing with time:
\begin{eqnarray}
\frac{d\mathcal{F}_d}{dt} & =&\sum_n \frac{d\phi_n}{dt} \times 
             \frac{\partial\mathcal{F}_d}{\partial\phi_n}\\
    &=&- \sum_n \left(\frac{d\phi_n}{dt} \right)^2 \leq 0. 
	\end{eqnarray}
Therefore, the free energy is guaranteed to decrease toward a minimum. 
Furthermore, the $\mathcal{L}^2$-norm of the rate of change of $\phi$ is 
proportional to the rate of change of the discretized energy. This means 
that if the interface is moving, then its squared velocity is proportional to the 
rate of change of the free energy. In other words, any motion of the interface 
(whatever are the discretization errors) comes with a decrease in  the free energy.
This differs qualitatively from the expression of Eq.~(\ref{disc_from_PDE}) which
\textit{a priori} has no reason to derive exactly from any energy and where
the interface can have a motion proportional to discretization errors, as
was indeed found in our early investigations of this model \cite{Mellenthin07}.

\section{Simulations at constant volume of solid}

The discretized equation of motion for the phase field has the
structure
\begin{equation}
\frac{d\phi_n}{dt} = A_n + B_n u_n,
\end{equation}
where a one-dimensional example is considered for simplicity, $A_n$
and $B_n$ are coefficients that depend on the local configuration of
phase and orientation fields, and $u_n$ is the discretized temperature
field. For a constant but time-dependent temperature, $u_n\equiv u(t)$,
we have
\begin{equation}
\frac{d}{dt} \sum_n \phi_n = \sum_n A_n + u(t) \sum_n B_n.
\end{equation}
The left hand side can be interpreted as the change in the total volume
of solid. Clearly, $\sum_n \phi_n$ can be kept constant by the following
procedure: (i) evaluate the coefficients $A_n$ and $B_n$, (ii) calculate
the temperature $u=- \sum_n A_n /(\sum_n B_n)$ which makes the left hand
side vanish, and (iii) timestep the equations with this value of the
temperature. After a short time, the temperature converges to a constant
value, which corresponds to the equilibrium temperature for the chosen
total volume of solid.

\end{document}